\documentclass[reprint,a4paper,australian,amsmath,aps,prl,twocolumn,superscriptaddress,preprintnumbers,noshowpacs]{revtex4-1}
\usepackage{times}
\usepackage{graphicx}
\usepackage{amssymb}
\usepackage{amsmath}
\usepackage{mathrsfs}
\usepackage[utf8]{inputenc}
\usepackage{xcolor}

\newcommand{\be}{\begin{equation}}
\newcommand{\ee}{\end{equation}}

\newcommand{\CSSM}{Special Research Centre for the Subatomic Structure
  of Matter (CSSM),\\Department of Physics, University of
  Adelaide, Adelaide, South Australia 5005, Australia} 

\newcommand{\CoEPP}{ARC Centre of Excellence for Particle Physics at
  the Terascale (CoEPP),\\Department of Physics, University
  of Adelaide, Adelaide, South Australia 5005, Australia} 

\newcommand{\NCI}{National Computational Infrastructure
  (NCI),\\Australian National University, Canberra, Australian Capital Territory
  0200, Australia}

\begin{document}

\preprint{ADP-16-49/T1005}

\title{Light-quark contributions to the magnetic form factor of the $\mathbf{\Lambda(1405)}$}

\author{Jonathan M. M. Hall} 
\author{Waseem Kamleh} 
\author{Derek B. Leinweber}
\affiliation{\CSSM}
\author{Benjamin J. Menadue}
\affiliation{\CSSM}
\affiliation{\NCI}
\author{Benjamin J. Owen}
\affiliation{\CSSM}
\author{Anthony W. Thomas} 
\affiliation{\CSSM}
\affiliation{\CoEPP}


\begin{abstract}
In a recent study of the $\Lambda(1405)$, the suppression of the
strange-quark contribution to the magnetic form factor was interpreted
as the discovery of a dominant antikaon-nucleon composition for this
low-lying state.  We confirm this result by calculating the light $u$-
and $d$-quark contributions to the $\Lambda(1405)$ magnetic form
factor in lattice QCD in order to determine the extent to which their
contributions support this exotic molecular description. Drawing on
the recent graded-symmetry approach for the flavor-singlet components
of the $\Lambda(1405)$, the separation of connected and disconnected
contributions is performed in both the flavor-octet and singlet
representations.
The relationship between light-quark contributions to the
$\Lambda(1405)$ magnetic form factor and the connected contributions
of the nucleon magnetic form factors is established and compared with
lattice calculations of the same quantities, confirming the $\overline{K}N$ molecular
structure of the $\Lambda(1405)$ in lattice QCD.
\end{abstract}

\maketitle


Resolving and understanding the internal structure of hadronic excited
states is an important contemporary problem in the field of
nonperturbative QCD.  While lattice QCD simulation methods are
increasingly able to probe the chiral regime of ground state
observables with unprecedented accuracy
\cite{Aoki:2008sm,Collins:2011mk,Green:2015wqa,Sufian:2016pex},
the resolution of excited-baryon form factors is still at a very early stage
\cite{Owen:2013pfa, Roberts:2013ipa, Roberts:2013oea, Owen:2014txa, Menadue:2013kfi, Hall:2014uca}.

Interest in the $\Lambda(1405)$ resonance has continued unabated for more than 50 years
\cite{Dalitz:1960du, Dalitz:1967fp, Veit:1984an, Veit:1984jr, Leinweber:1989hh, Kaiser:1995eg, Lage:2009zv, Kaiser1997, Oset1998, Oller2001, GarciaRecio:2002td, Jido:2002yz, Jido:2003cb, Hyodo2004, Magas:2005vu, Geng:2007hz, Doring:2010rd, Ikeda2011, Guo2013, Mai2013, Molina:2015uqp, Liu:2016wxq, Mai:2012dt, MartinezTorres:2012yi, Miyagawa:2012xz, Sekihara:2013sma, Xie:2013wfa, Oller:2013zda, Hall:2014uca, Hall:2014gqa, Menadue:2011pd, Engel:2012qp, Moriya:2013eb, Roca:2013av, Hall:2015cua} 
because of its unusually low mass -- lower even than the corresponding mass of the negative parity nucleon, despite containing
a heavier strange quark.
The unexpected position of the $\Lambda(1405)$ in the spectrum 
has been explored in several studies,
which typically indicate a significant contribution from a $\overline{K} N$ bound state
\cite{Dalitz:1960du, Dalitz:1967fp, Veit:1984an, Veit:1984jr, Leinweber:1989hh, Kaiser:1995eg,Lage:2009zv, Kaiser1997, Oset1998, Oller2001, GarciaRecio:2002td, Jido:2002yz, Jido:2003cb, Hyodo2004, Magas:2005vu, Geng:2007hz, Doring:2010rd, Ikeda2011, Guo2013, Mai2013, Molina:2015uqp, Liu:2016wxq, Mai:2012dt, MartinezTorres:2012yi, Miyagawa:2012xz, Sekihara:2013sma, Xie:2013wfa, Oller:2013zda, Hall:2014uca, Hall:2014gqa}.
The $\pi\Sigma$ channel also plays a nontrivial role.  It is now widely agreed that there is a
two-pole structure in this resonance region
\cite{Kaiser1997,Oset1998,Oller2001,GarciaRecio:2002td,Jido:2002yz,Jido:2003cb,Hyodo2004,Magas:2005vu,Geng:2007hz,Doring:2010rd,Ikeda2011,Guo2013,Mai2013,Molina:2015uqp,Liu:2016wxq} 
stemming from attractive interactions in both the $\pi\Sigma$ and $\overline{K}N$ channels.
In making contact with results from lattice QCD \cite{Menadue:2011pd,Engel:2012qp,Hall:2014uca}, a
description of the $\Lambda(1405)$ over a range of quark masses has been developed
\cite{Hall:2013qba,Hall:2014uca,Liu:2016wxq}, bridging constituent-quark ideas at heavy quark
masses and the molecular $\overline{K} N$ dominance of the $\Lambda(1405)$ at light quark masses.

A recent lattice QCD study of the $\Lambda(1405)$ reported evidence of
a molecular $\overline{K} N$ structure~\cite{Hall:2014uca}.  There, the
role of the strange quark was paramount in signaling the presence of a
dominant $\overline{K} N$ structure.  At heavier quark masses
approaching the strange quark mass, the three quark flavors ($u$,
$d$, $s$) are found to make approximately equal contributions to the
magnetic form factor when their charges are set to unity.  The
underlying flavor symmetry is manifest. However, as the $u$ and $d$
quarks become light, flavor symmetry in the quark contributions to the
magnetic form factor is found to be badly broken, and the strange-quark
contribution drops by an order of magnitude from its maximum 
to a nearly vanishing value at the smallest quark mass.

This feature has a simple explanation in terms of a $\overline{K} N$
molecule.  The strange quark is confined in a spin-$0$ kaon in a
relative $S$ wave about the nucleon, implying a net absence of angular
momentum. Hence, the strange quark cannot contribute to the magnetic
form factor of a $\Lambda(1405)$ composed as a molecular $\overline{K}
N$ bound state.

In this Letter we focus on the light-quark sector of the magnetic form
factor of the $\Lambda(1405)$ in lattice QCD.  Until now, it has
received little attention. Nevertheless, it is a vital piece of
information in the quest to confirm whether the lattice QCD value
supports the $\overline{K} N$ molecular description, and is complementary
to the strange sector analysis of Ref.~\cite{Hall:2014uca}.

The analysis of the light-quark sector is not straightforward.  
Careful attention must be
given to what has (and has not) been included in the lattice QCD calculation.
In particular, the calculations so far~\cite{Hall:2014uca} omit photon couplings to
quark--antiquark loops in the vacuum, which in turn interact with the connected quarks via gluon
exchange.
These so-called disconnected loop contributions are unlikely to be determined in the near
future because of the difficulty they present in numerical simulations of baryon excited states.
As the loop is correlated with the quarks carrying the quantum numbers of the state only via gluon
exchange, resolving a nontrivial signal requires high statistics and innovative methods.  While
there has been recent success in isolating the relevant disconnected-loop contributions in
ground-state baryon matrix elements \cite{Green:2015wqa,Sufian:2016pex}, challenges in isolating
baryon excitations in lattice QCD %
\cite{Mahbub:2010rm,Menadue:2011pd,Edwards:2011jj,Mahbub:2012ri,Lang:2012db,Mahbub:2013ala,Morningstar:2013bda,Alexandrou:2014mka,Owen:2015fra,Hall:2013qba,Hall:2014uca,Liu:2015ktc,Leinweber:2015kyz,Liu:2016uzk,Lang:2016hnn}
render the resolution of disconnected contributions elusive.

Here, we draw on partially-quenched chiral effective field theory
\cite{Labrenz:1996jy,Labrenz:1993wt,Savage:2001dy,Chen:2001yi,Leinweber:2002qb,Beane:2002vq,Detmold:2006vu,Hall:2013dva,Shanahan:2013apa,Hall:2015cua} 
to understand the relative weight of
these disconnected contributions to the form factors in QCD.
With this insight, one can test quantitatively whether the light-quark contribution to the magnetic
form factor of the $\Lambda(1405)$, calculated in lattice QCD, is consistent with a molecular
$\overline{K} N$ description of the internal structure. 

In the $\overline{K} N$ picture, the spin-$0$ kaon is in a relative $S$ wave about
the nucleon.  Therefore the light-quark contributions to the magnetic form factor of the
$\Lambda(1405)$ have their origin solely in the magnetic form factors of the nucleon.  As the couplings for
$\Lambda^* \to K^-\, p$ and $\Lambda^* \to \overline{K}^0 \, n$ are equal, the light sector
contribution is related to an average of $n$ and $p$ magnetic form factors in full QCD.

To explore this in further detail, consider the following simple model for the $\Lambda(1405)$
\begin{equation}
|\, \Lambda^* \rangle =  \frac{1}{\sqrt{2}} \left ( \, |\, K^- p \rangle + |\, \overline{K}^0 n
\rangle \, \right ) \, .
\label{eq:simpleModel}
\end{equation}
In full QCD (with disconnected sea-quark loop contributions included), 
the form of the quark sector contributions to the light-quark magnetic
form factor $\mu_q(Q^2)$ is simple:
\begin{eqnarray}
\langle \Lambda^* \,|\, \hat\mu_q \,|\, \Lambda^* \rangle 
&=& \frac{1}{2} \langle K^- p \,|\, \hat\mu_q \,|\, K^- p \rangle + \frac{1}{2} \langle \overline{K}^0 n
\,|\, \hat\mu_q \,|\, \overline{K}^0 n \rangle \, , \nonumber \\
&=& \frac{1}{2} \langle p \,|\, \hat\mu_q \,|\, p \rangle + \frac{1}{2} \langle n
\,|\, \hat\mu_q \,|\,  n \rangle \, .
\label{eq:nucleonRelation}
\end{eqnarray}
Here the zero spin and relative $S$ wave orbital angular
momentum of the kaon about the nucleon has been taken into account.  As $m_u =
m_d$ in the lattice QCD simulations~\cite{Hall:2014uca}, we consider
the charge-symmetric limit of the nucleon magnetic form factors.
Since the disconnected sea-quark loop contributions to the magnetic
form factor are not accessible for the $\Lambda(1405)$, we neglect
them consistently throughout our comparison to the nucleon magnetic
form factors.

To make the charge symmetry manifest in our results~\cite{Leinweber:1995ie}, we work with single
quarks of unit charge~\cite{Leinweber:1990dv,Leinweber:1999nf,Boinepalli:2006xd}, and define the operator $\hat
\mu_q$, omitting the electric charge factors of $2/3$ and $-1/3.$
The doubly- and singly-represented quark sector contributions to the nucleon form factors are
defined as $u_p = d_n$ and $d_p = u_n$, respectively, where the subscripts indicate the baryon in
which the quark resides.
The connected contributions to the nucleon form factors in the charge-symmetric limit are then
\begin{align}
\langle p \,|\, \hat\mu_u \,|\, p \rangle &= 2\, u_p \, , \qquad
\langle n \,|\, \hat\mu_u \,|\, n \rangle  =  1\, u_n = 1\, d_p \, , 
\label{eq:uInNucleon} \\
\langle p \,|\, \hat\mu_d \,|\, p \rangle &= 1\, d_p \, , \qquad
\langle n \,|\, \hat\mu_d \,|\, n \rangle  =  2\, d_n = 2\, u_p\, , 
\label{eq:dInNucleon} 
\end{align}
where the numerical factor counts the quarks.
These matrix elements are readily calculated in lattice QCD via the methods introduced in
\cite{Leinweber:1990dv}.

Returning now to the $\overline{K} N$ picture, Eq.~(\ref{eq:nucleonRelation}) yields a $u$-quark contribution to the 
$\Lambda(1405)$ magnetic form factor given by
\begin{equation}
\langle \Lambda^* \,|\, \hat\mu_u \,|\, \Lambda^* \rangle 
= \frac{1}{2} \left ( \, 2\, u_p + d_p \, \right ) \, ,
\label{eq:uInLambda}
\end{equation}
where a proton labeling has been used for $u_n = d_p$.  Similarly, the $d$-quark contribution is
\begin{equation}
\langle \Lambda^* \,|\, \hat\mu_d \,|\, \Lambda^* \rangle 
= \frac{1}{2} \left ( \, d_p + 2\, u_p \, \right ) \, .
\label{eq:dInLambda}
\end{equation}
Thus the isospin-symmetry of the
$\Lambda(1405)$
%
%
is manifest with a light-quark contribution of
\begin{equation}
\langle \Lambda^* \,|\, \hat\mu_\ell \,|\, \Lambda^* \rangle 
= u_{\Lambda^*} = d_{\Lambda^*} 
= \frac{1}{2} \left ( \, 2\, u_p + d_p \, \right ) \, .
\label{eq:lInLambda}
\end{equation}

While we have been careful to omit disconnected sea-quark loop contributions to the nucleon form
factors, our simple $\overline{K} N$ model includes an implicit disconnected contribution that
has not been included in the lattice QCD calculation of the $\Lambda(1405)$ magnetic form factor.
We now identify that contribution, calculate it, and remove it from Eq.~(\ref{eq:lInLambda}).
\begin{figure}[t]
\begin{center}
\includegraphics[width=0.50\columnwidth,angle=0]{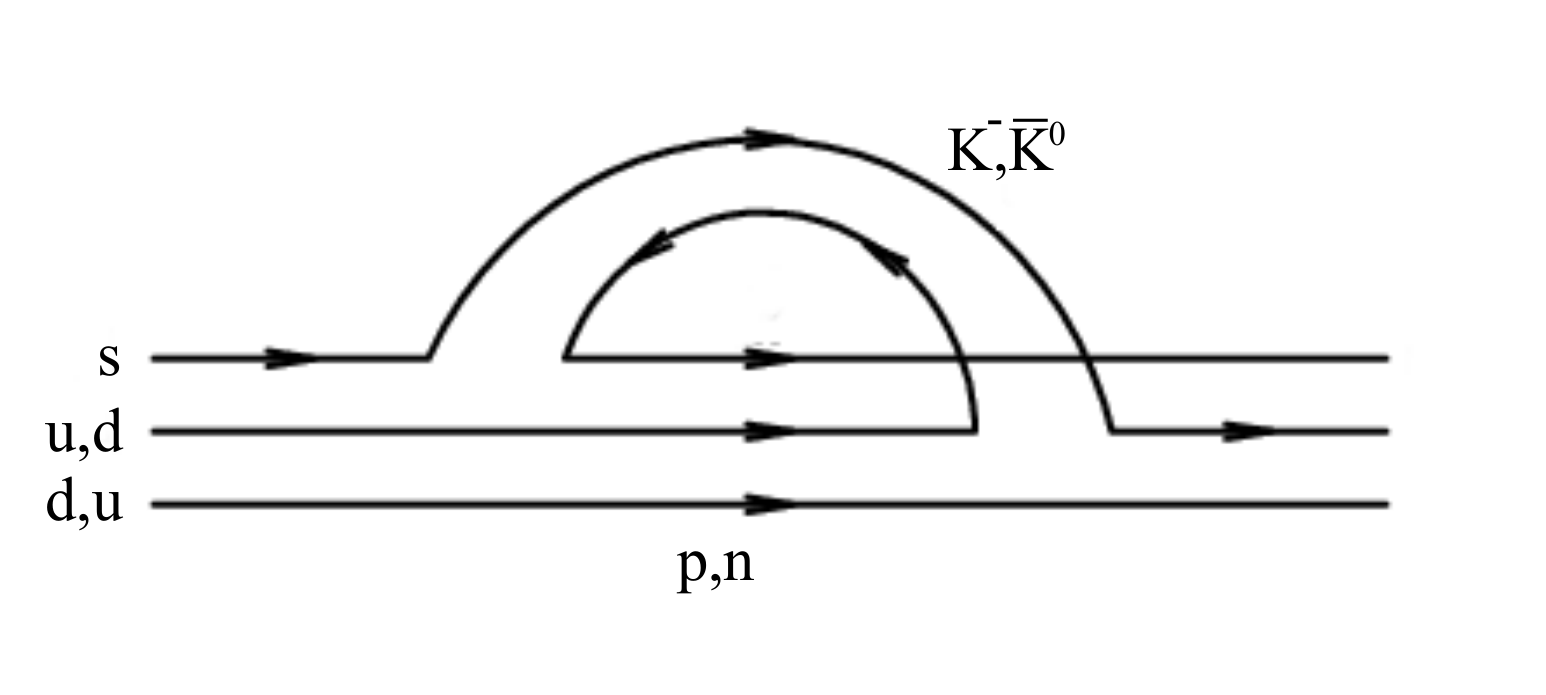}\hspace{-5mm}
\includegraphics[width=0.50\columnwidth,angle=0]{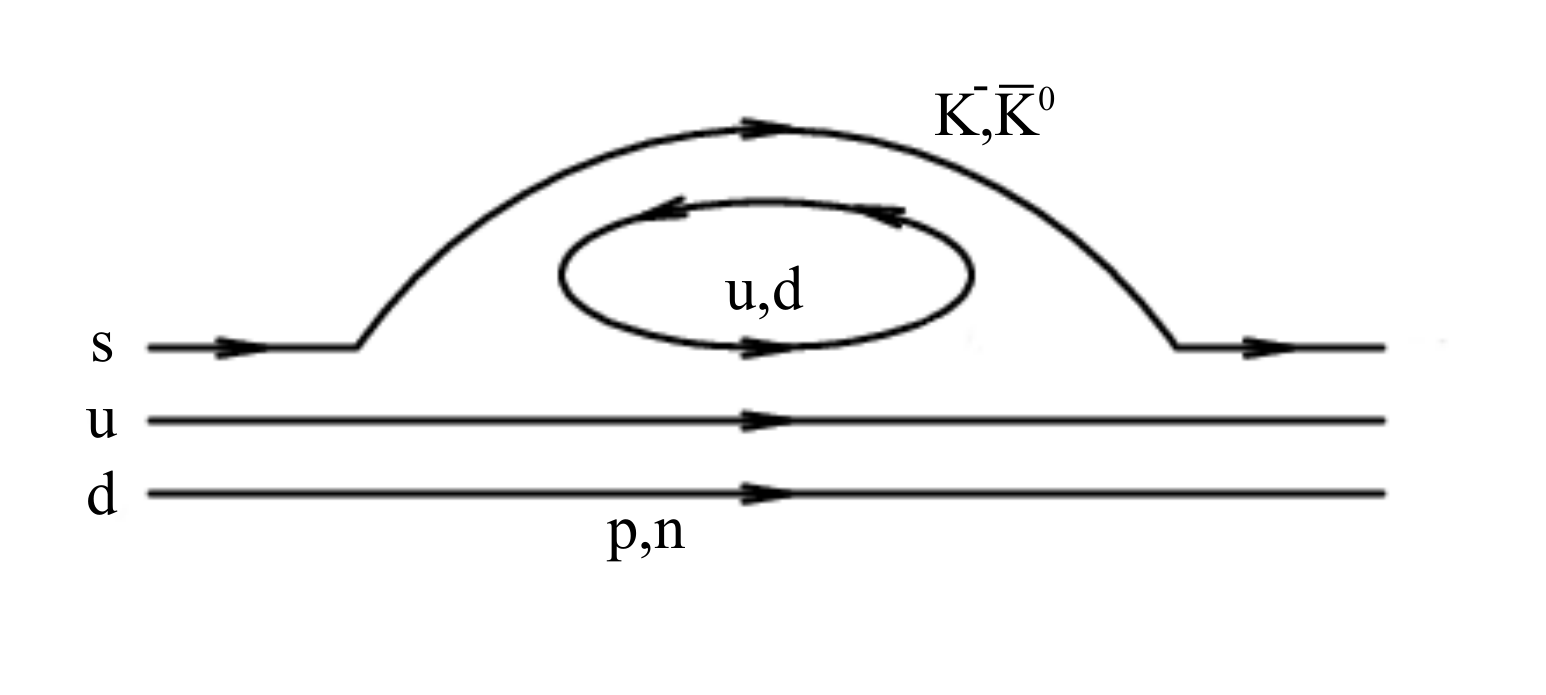}
\vspace*{-8pt}
\caption{The leading-order loop contributions from the process: $\Lambda(1405)\rightarrow\overline{K}N$. 
\vspace*{-12pt}
\label{fig:split}}
\end{center}
\end{figure}

{}Figure~\ref{fig:split} illustrates the connected and disconnected
$\overline K N$ loop contributions to the two-point function governing
the mass of the $\Lambda(1405)$ in full QCD.  As the lattice
calculations are performed on $2+1$ flavor dynamical fermion gauge
field configurations, both of these diagrams are included.

\begin{figure}[t]
\begin{center}
\includegraphics[width=0.5\columnwidth,angle=0]{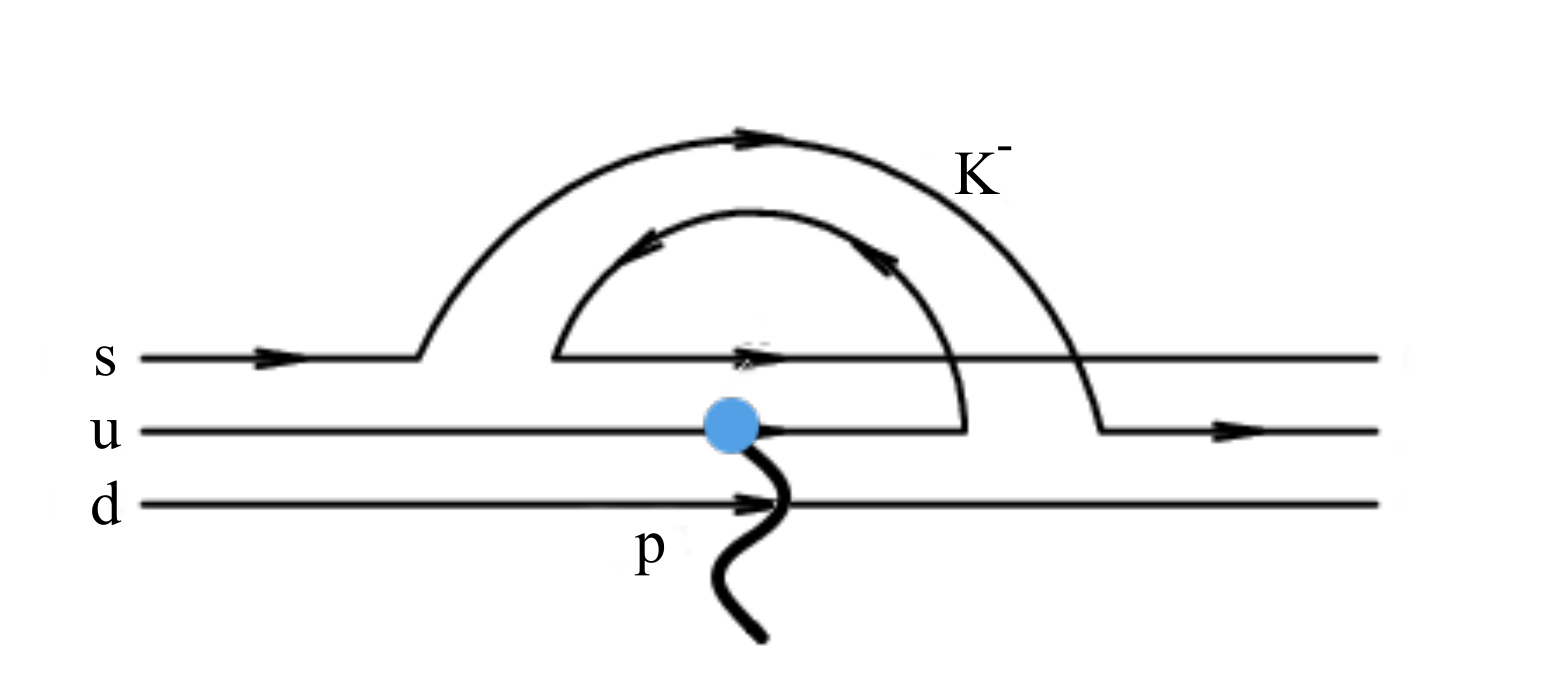}\hspace{-5mm}
\includegraphics[width=0.5\columnwidth,angle=0]{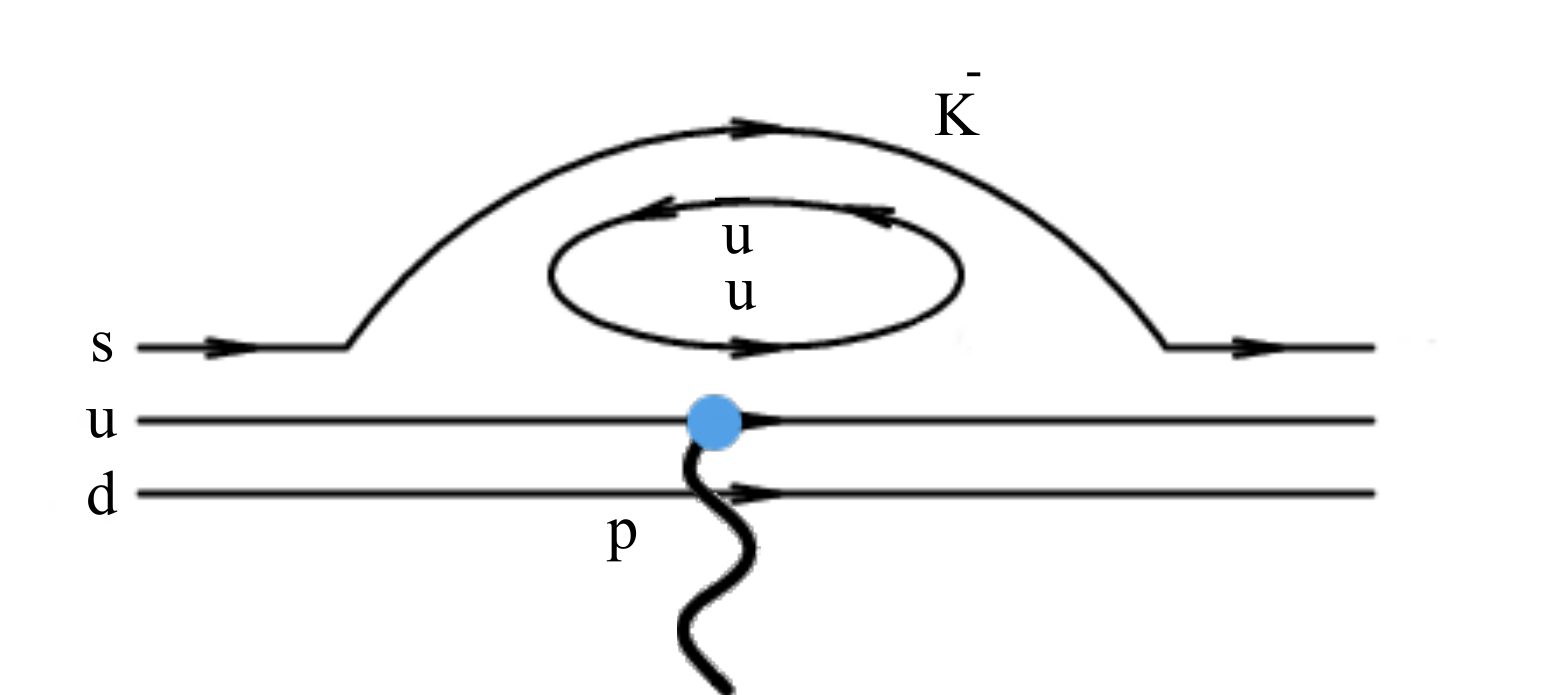}
\includegraphics[width=0.5\columnwidth,angle=0]{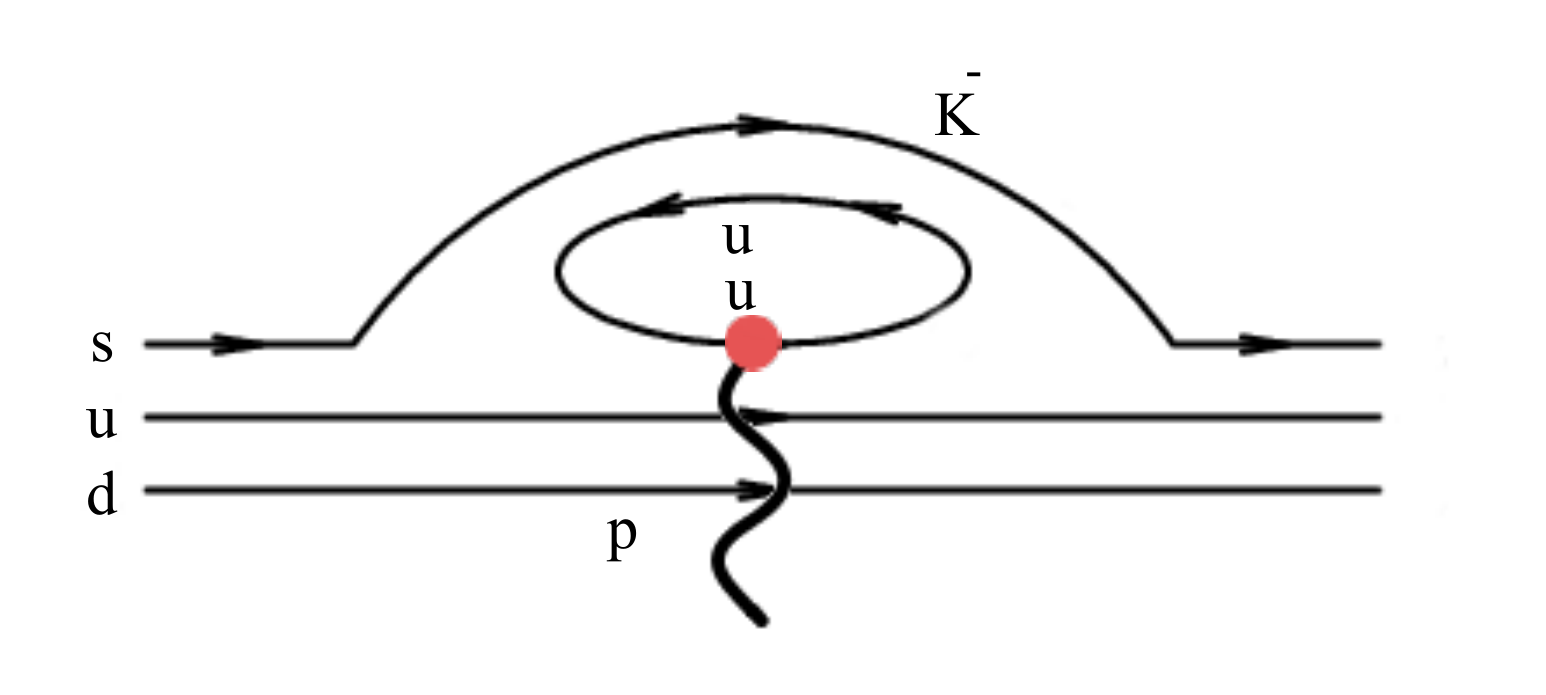}
\vspace*{-8pt}
\caption{(color online). The quark flow diagrams for the process $\Lambda(1405) \rightarrow K^-\, p$ can be
  decomposed into a completely-connected part and two parts involving disconnected sea-quark loop
  contributions.  The upper-left completely-connected diagram and the upper-right diagram are
  included in the lattice QCD calculations as the photon couples to a quark flowing in a connected
  manner from the source to the sink.  The case where a photon couples to a disconnected sea quark
  loop, illustrated in the lower diagram, is not included in the lattice QCD calculations of
  $\langle \Lambda^* \,|\, \hat\mu_\ell \,|\, \Lambda^* \rangle$.
\vspace*{-12pt}
\label{fig:dcomp}}
\end{center}
\end{figure}

The difficulty with sea-quark loop contributions to the magnetic form
factor of the $\Lambda(1405)$ is illustrated in Fig.~\ref{fig:dcomp},
where $u_p$ is considered.  Recalling that photon couplings to the
spin-$0$ kaon in a relative $S$ wave about the nucleon do not contribute
to the magnetic form factor of the $\Lambda(1405)$, the focus is on
the nucleon couplings.  In the upper quark-flow diagrams of
Fig.~\ref{fig:dcomp}, the photon couples to $u$ quarks flowing from
source to sink. These connected insertions of the photon current are
included in the lattice QCD calculations of $\langle \Lambda^* \,|\,
\hat\mu_\ell \,|\, \Lambda^* \rangle$.

However, the coupling of the photon to the disconnected sea quark loop, illustrated in the
lower diagram of Fig.~\ref{fig:dcomp}, is not included.  As the upper-right and lower diagrams
contribute with equal weight, half of the disconnected sector is included, and half is omitted.
The task that remains is to understand the relative contributions of the fully-connected diagram
and those involving a disconnected sea-quark loop.  Thus, we return our attention to
Fig.~\ref{fig:split}.

To determine the relative weight of the couplings between the connected and disconnected diagrams
of Fig.~\ref{fig:split}, 
we draw upon partially-quenched chiral perturbation theory
\cite{Labrenz:1996jy,Labrenz:1993wt,Savage:2001dy,Chen:2001yi,Leinweber:2002qb,Beane:2002vq,Detmold:2006vu,Hall:2013dva,Shanahan:2013apa,Hall:2015cua}.
For $\Lambda$ baryons composed of three quark flavors, the graded
symmetry approach
\cite{Labrenz:1996jy,Labrenz:1993wt,Savage:2001dy,Chen:2001yi,Hall:2015cua}
is preferred over the diagrammatic
approach~\cite{Leinweber:2002qb,Hall:2013dva}.  The graded symmetry
approach extends standard chiral perturbation theory by introducing
commuting ghost field counterparts ($\tilde{u}$, $\tilde{d}$,
$\tilde{s}$) to the three quarks ($u$, $d$, $s$). As the ghost-quark
fields only enter the interaction through the disconnected loops, they
allow the decomposition of the quark flow diagrams into connected and
disconnected parts. This can be seen in Fig.~\ref{fig:split}, where
the completely connected diagram contains only valence quarks and the
disconnected loop diagram allows sea quarks to contribute to the
amplitude. The contributions from the disconnected diagram are
isolated by extracting the ghost meson-baryon contribution to each
vertex in the diagram.

In the $SU(3)$-flavor limit, the $\Lambda(1405)$ is identified as the low-lying flavor-singlet
baryon.  However, as one approaches the physical regime, significant mixing with octet-flavor
symmetry is encountered~\cite{Menadue:2011pd,Hall:2014uca}.  Therefore one needs to consider both
flavor-octet and flavor-singlet couplings for $\Lambda^* \rightarrow \overline{K} N$.  In
addressing the latter, we draw upon the recently developed graded-symmetry approach for singlet
baryons, augmenting the standard octet-baryon Lagrangian with the necessary additional terms
\cite{Hall:2015cua}.

First, we consider the contributions to the singlet component of the $\Lambda(1405)$, denoted
$\Lambda^{\prime *}$, where the prime indicates that a singlet representation is taken, and the
star indicates that the resonance has odd parity.  In the case of the process $\Lambda^{\prime
  *}\rightarrow K^-\, p$, the relevant ghost term in the Lagrangian takes the form
\begin{equation}
- g_s\, \sqrt{\frac{2}{3}}\; \overline{\Lambda}^{\prime *}\, \tilde{K}^-\,
\tilde{\Lambda}^+_{p,\tilde{u}} \, ,
\end{equation}
where $g_s$ is taken to be the coupling of the singlet to octet-octet process $\Lambda^{\prime *}
\rightarrow \pi_0\, \Sigma_0$.  Here, we follow the notation of Ref.~\cite{Hall:2015cua}:
$\tilde{K}^-$ is composed of a strange quark ($s$) and a ghost anti-up quark
($\overline{\tilde{u}}$) and $\tilde{\Lambda}^+_{p,\tilde{u}}$ represents a proton-like particle
composed of $\tilde{u}ud$, with the normal quarks in an anti-symmetric formation.  The factor
$\sqrt{2/3}$ is derived from the $SU(3|3)$ symmetry relations that govern the Lagrangian.  
With reference to the full QCD amplitude,
\begin{equation}
g_s\, \overline{\Lambda}^{\prime *}\, K^-\, p \, ,
\end{equation}
the relative weights of the diagrams in Fig.~\ref{fig:split} can be
resolved.  As a consequence of flavor symmetry, the connected diagram
has weight $(1/3)\, g_s^2$ and the disconnected diagram has weight
$(2/3)\, g_s^2$. Similar results are found for $\Lambda^{\prime *}
\rightarrow \overline{K}^0\, n$, where a $d$ quark participates in the
loop in full QCD, such that a comparison with the partially quenched
term resolves the same weightings as above.

Significant flavor-symmetry breaking in the physical quark-mass regime admits an important
flavor-octet symmetry in the structure of the $\Lambda(1405)$.  Thus, one must also consider
octet-to-octet meson and baryon contributions.  Upon partial quenching,
the corresponding couplings derived are
\begin{equation}
\frac{\sqrt{2}(D+3F)}{3}\;\left\{ \, \begin{matrix} \overline{\Lambda}^{*}\, \tilde{K}^-\,\tilde{\Lambda}^+_{p,\tilde{u}} \\ \overline{\Lambda}^{*}\, \overline{\tilde{K}}^0\,\tilde{\Lambda}^0_{n,\tilde{d}} \end{matrix} \, \right\}
\end{equation}
for the $u$- and $d$-quark loops, respectively.
In full QCD, both the $\Lambda^* \to K^-\, p$ and $\Lambda^* \to \overline{K}^0\, n$ channels have
the coupling $-(D+3F)/\sqrt{3}$.  Thus, one observes the same ratio of $\sqrt{2/3}$ between the
disconnected sea-quark loop component couplings and the full QCD couplings.  Again, the connected
diagram holds a weight of $1/3$ and the disconnected diagram holds a weight of $2/3$ of the full QCD process.
As the split between connected and disconnected components is the same for the different flavor
representations, the calculation of the partially quenched value of the magnetic form factor
is straightforward.

The ratio between the connected and disconnected weights determines the extent to which the full
QCD magnetic form factor is changed on the lattice due to the omission of photon couplings to the
disconnected sea-quark loops.  Consider, for example, the $u$-quark contribution in the proton,
$u_p$, where the $K^-\, p$ intermediate state contains a disconnected $u$-quark contribution.
While one-third of the result is preserved in the connected contribution, only half of the
remaining two-thirds involving disconnected contributions is preserved.  Thus, one can obtain the
$u$-quark contributions to the proton that are included in the lattice QCD calculations by
subtracting off $1/2\times 2/3 = 1/3$ of the full QCD contribution.
The $u$-quark contribution to the neutron, $u_n$, is fully included in the lattice QCD
calculation as, in the $\Lambda^{\prime *} \rightarrow \overline{K}^0\, n$ channel, the disconnected
quark-loop flavor is a $d$ quark, not a $u$ quark, so no adjustment is required.
In summary, 
\begin{widetext}
\begin{equation}
\langle \Lambda^* \,|\, \hat\mu^{\rm conn}_u \,|\, \Lambda^* \rangle 
= \frac{1}{2}\, \left (
\langle K^- p \,|\, \hat\mu_u \,|\, K^- p \rangle 
- \frac{1}{2}\, \frac{2}{3} \, \langle K^- p \,|\, \hat\mu_u \,|\, K^- p \rangle
\right ) 
+ \frac{1}{2}\, \langle \overline{K}^0 n \,|\, \hat\mu_u \,|\, \overline{K}^0 n \rangle \, 
%
= \frac{1}{2}\, \left ( 2 u_p - \frac{2}{3} u_p  + u_n \right )\, .
\label{eqn:muu}
\end{equation}
\end{widetext}
The first two terms in the leading parentheses of Eq.~(\ref{eqn:muu}) represent the connected
$u$-quark contribution from the proton component within the $\Lambda(1405)$.  The first term
provides the full QCD contribution while the second term subtracts half of the weight of the
disconnected sea-quark loop associated with photon couplings to the disconnected loop. 
Similarly, for the $d$-quark contribution, we obtain a value of $\frac{1}{2}( 2 d_n - \frac{2}{3} d_n  + d_p)$, 
%
and under charge symmetry, the two light quark contributions become equal,
\begin{equation}
\langle \Lambda^* \,|\, \hat\mu^{\rm conn}_\ell \,|\, \Lambda^* \rangle 
= \frac{1}{2}\, \left ( 2 u_p - \frac{2}{3} u_p  + u_n \right )\, .
\label{eqn:finalModel}
\end{equation}


To test the $\overline{K} N$ model prediction of Eq.~(\ref{eqn:finalModel}), we draw on the same set
of configurations explored in Ref.~\cite{Hall:2014uca}. where the left-hand side of the equation,
$\langle \Lambda^* \,|\, \hat\mu^{\rm conn}_\ell \,|\, \Lambda^* \rangle$, was calculated.
These calculations are based on the $32^3 \times 64$ full-QCD
ensembles created by the PACS-CS collaboration~\cite{Aoki:2008sm},
made available through the International Lattice Data Grid (ILDG)
\cite{Beckett:2009cb}. The ensembles provide a lattice volume of
$(2.9\ \mbox{fm})^3$ with five different masses for the light $u$ and
$d$ quarks, and constant strange-quark simulation parameters.  We
simulate the valence strange quark with a hopping parameter of
$\kappa_s = 0.13665$, reproducing the correct kaon mass in the
physical limit \cite{Menadue:2012kc}.  We use the squared pion mass as
a renormalization group invariant measure of the quark mass.  The
scale is set via the Sommer parameter \cite{Sommer:1993ce} with $r_0 =
0.492$ fm \cite{Aoki:2008sm}.
The nucleon magnetic form factors are determined on these lattices using the methods introduced in
Ref.~\cite{Leinweber:1990dv} and refined in Ref.~\cite{Boinepalli:2006xd}, providing values of $u_p = 1.216(17)\;\mu_N$ and $u_n = -0.366(19)\;\mu_N$ at the lightest pion mass. Results are reported
for the lowest nontrivial momentum transfer of $Q^2 \simeq 0.16$ GeV${}^2/c^2$.

\begin{figure}[t]%
\centering%
\includegraphics[height=0.9\hsize,angle=90]{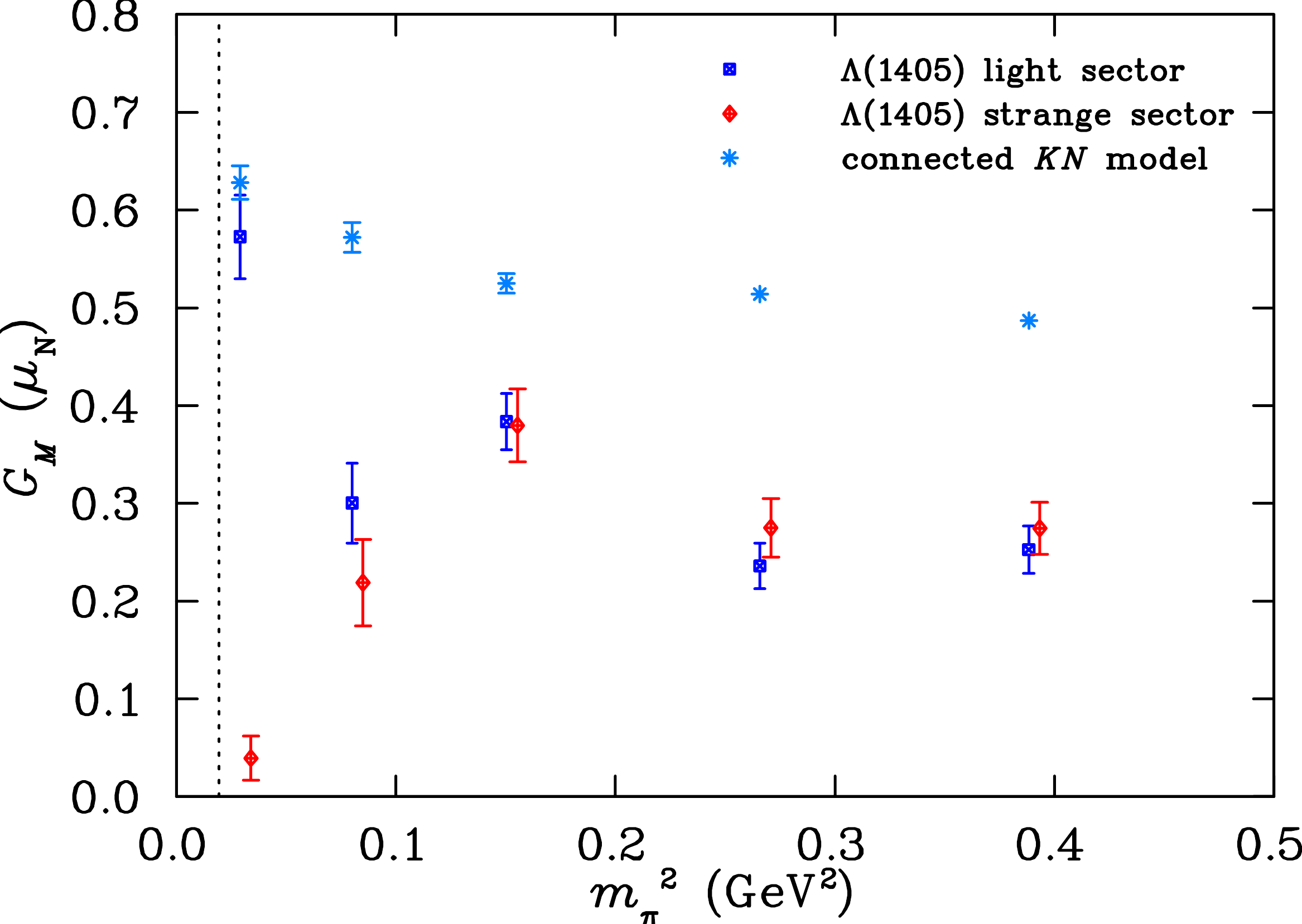}%
\vspace{-1mm}%
\caption{\footnotesize{The light ($u$ or $d$) and strange quark contributions to the magnetic form
    factor of the $\Lambda(1405)$ at $Q^2 \simeq 0.16$ GeV${}^2/c^2$ from Ref.~\cite{Hall:2014uca}
    are presented as a function of the light $u$- and $d$-quark masses, indicated by the squared
    pion mass, $m_\pi^2$.  Sector contributions are for single quarks of unit charge.  
The lattice calculations are compared to the predictions of the connected $\overline{K} N$ model developed
    herein and summarized in Eq.~(\ref{eqn:finalModel}).
    The vertical dashed line indicates the physical pion mass.
    The strange form factor results are offset a small amount from the light sector in the
    $m_\pi^2$-axis for clarity.}%
\label{fig:mpisq}}%
\end{figure}

Lattice QCD results from Ref.~\cite{Hall:2014uca} for the light- and
strange-quark magnetic form factors of the $\Lambda(1405)$ are plotted
as a function of pion mass in Fig.~\ref{fig:mpisq}.  As mentioned
earlier, the flavor symmetry present at heavy quark masses is broken
as the $u$ and $d$ masses approach the physical point, where the
strange magnetic form factor drops to nearly zero.  The light quark
sector contribution differs significantly from the molecular
$\overline{K} N$ model prediction until the lightest quark mass is
reached.  At this point, the direct matrix element calculation, $\langle
\Lambda^* \,|\, \hat\mu^{\rm conn}_\ell \,|\, \Lambda^* \rangle$ of
Ref.~\cite{Hall:2014uca}, ``$\Lambda(1405)$ light sector'' in
Fig.~\ref{fig:mpisq}, agrees with the prediction of the ``connected
$\overline{K} N$ model'' developed here and summarized in
Eq.~(\ref{eqn:finalModel}).  This agreement confirms that the
$\Lambda(1405)$ observed in lattice QCD at quark masses resembling
those of Nature is dominated by a molecular $\overline{K} N$
structure.  At the lightest pion mass, the light-quark magnetic form
factor of the $\Lambda(1405)$ is \cite{Hall:2014uca}
\begin{equation}
\langle \Lambda^* \,|\, \hat\mu^{\rm conn}_\ell(Q^2) \,|\, \Lambda^* \rangle  = 0.58(5)\ \mu_N \, ,
\end{equation}
at $Q^2 \simeq 0.16$ GeV${}^2/c^2$.  The connected $\overline{K} N$ model of
Eq.~(\ref{eqn:finalModel}) predicts
\begin{equation}
\langle \Lambda^* \,|\, \hat\mu^{\rm conn}_\ell(Q^2) \,|\, \Lambda^* \rangle 
= 0.63(2)\ \mu_N \, .
\label{eqn:finalPrediction}
\end{equation}
It is important to note that the shift in the prediction due to the omission of photon couplings
to the disconnected sea-quark loop is significant.  In the case where such couplings are included,
the prediction of the $\overline{K} N$ model is significantly larger at $\langle \Lambda^* \,|\,
\hat\mu_\ell \,|\, \Lambda^* \rangle = (2\, u_p + u_n) / 2 = 1.03(2)\ \mu_N$. Thus, it is important
for the lattice community to continue to work towards a determination of these disconnected-loop
contributions, particularly for resonances where coupled channel dynamics play an important role.

The light-quark sector contributions to the magnetic form factor of
the $\Lambda(1405)$ calculated in lattice QCD \cite{Hall:2014uca} have
been examined in the context of a molecular $\overline{K} N$ model in
which the quark-flow connected contributions to the magnetic form
factor have been identified.  This enables a quantitative analysis of
the extent to which the light-quark contributions are consistent with
a molecular bound-state description.

Identification and removal of the quark-flow disconnected
contributions to the $\overline{K} N$ model have been made possible
via a recently developed graded-symmetry approach~\cite{Hall:2015cua}.
It is interesting to note that the relative contribution of connected
to disconnected contributions is in the ratio $1:2$ for both
flavor-singlet and flavor-octet representations of the $\Lambda$
baryon.  

Using new results for the magnetic form factors of the nucleon at a
near-physical quark mass of $m_\pi = 156$ MeV, the connected
$\overline{K} N$ model predicts a light-quark sector contribution to
the $\Lambda(1405)$ of $0.63(2)\ \mu_N$, which agrees remarkably well
with the direct calculation of $0.58(5)\ \mu_N$ from
Ref.~\cite{Hall:2014uca}. This confirms that the internal structure of
the $\Lambda(1405)$ is dominated by a $\overline{K} N$ molecule.

The $\Lambda(1405)$ observed in lattice QCD has significant overlap with
local three-quark operators and displays a dispersion relation
consistent with that of a single baryon. This implies that the
$\overline{K} N$ bound state is localised. Furthermore, it is striking
that the nucleon maintains its properties so well when bound.

Future work will focus on the isolation of nearby multi-particle scattering states in the finite
lattice volume, and explore their quark sector contributions to the magnetic form factors.  Using
the formalism developed in Ref.~\cite{Briceno:2015tza} one can then combine all the low-lying contributions
observed in lattice QCD and make contact with the full resonance structure.

\begin{acknowledgments}
{\bf Acknowledgements:} 
We thank the PACS-CS Collaboration for making their $2+1$ flavor configurations available and the
ongoing support of the ILDG.  This research was undertaken with the assistance of the University of
Adelaide's Phoenix cluster and resources at the NCI National Facility in Canberra, Australia.  NCI
resources were provided through the National Computational Merit Allocation Scheme, supported by
the Australian Government and the University of Adelaide Partner Share.  This research is supported
by the Australian Research Council through the ARC Centre of Excellence for Particle Physics at the
Terascale (CE110001104), and through Grants No.\ LE160100051, DP151103101 (A.W.T.), DP150103164,
DP120104627 and LE120100181 (D.B.L.).
\end{acknowledgments}


\bibliography{refs}

\end{document}